\documentclass[aps,prb,twocolumn,groupedaddress]{revtex4-1}
\usepackage[latin1]{inputenc}
\usepackage[T1]{fontenc}
\usepackage{bm}
\usepackage{graphicx}
\usepackage{amsmath}
\usepackage{amsfonts}
\usepackage{amssymb}
\usepackage{epstopdf}
\DeclareGraphicsExtensions{.pdf,.eps,.png,.jpg,.mps} 
\begin{document}

\title {Wave function engineering in quantum dot--ring nanostructures}
\author{El\.zbieta Zipper} 
\author{Marcin Kurpas}
\author{Maciej M. Ma\'{s}ka}
\email{maciek@phys.us.edu.pl}
\affiliation{Institute of Physics, University of Silesia, Uniwersytecka 4, 40-007 Katowice, Poland }

\begin{abstract}
Modern nanotechnology allows producing, depending on application, various quantum nanostructures with the desired properties. These properties are strongly influenced by the confinement potential which can be modified e.g., by electrical gating.
In this paper we analyze a nanostructure composed of a quantum dot surrounded by a quantum ring. We show that depending on the details of the confining potential the electron wave functions can be located in different parts of the structure. Since the properties of such a nanostructure strongly depend on the distribution of the wave functions, varying the applied gate voltage one can easily control them. In particular, we illustrate the high controllability of the nanostructure by demonstrating how its coherent, optical, and conducting properties can be drastically changed by a small modification of the confining potential.
\end{abstract}

\maketitle
\section{Introduction}\label{sec1}
Quantum nanostructures \cite{aws,hans} are frequently referred to as artificial atoms. Like the natural atoms they show a discrete spectrum of energy levels but they exhibit new physics which has no analogue in real atoms. In particular, the electronic properties of quantum nanostructures can be finely tuned adjusting structural parameters such as size and shape. The latter parameter is particularly important as its small variations can cause dramatic changes of the electronic properties.\cite{amasha}
Nowadays nanotechnology enables a precise control of structural parameters both at the fabrication stage of quantum nanostructures\cite{somaschini,shorubalko} as well as dynamically while operating the device, e.g., by electrostatic potential.\cite{amasha} Exceptionally high degree of tunability can be achieved in complex nanostructures composed of coupled elements, where the coupling constant is an additional parameter that can be controlled. A high tunability has been already demonstrated in systems of coupled quantum dots\cite{dqd} and quantum rings.\cite{dqr} 

In this paper we discuss properties of a similar system, namely a two-dimensional nanostructure in the form of a quantum dot (QD) surrounded by a quantum ring (QR) named afterwords a dot-ring nanostructure (DRN). Such a structure has recently been fabricated by droplet epitaxy.\cite{somaschini}
We show that by changing the parameters of the potential barrier $V_0(r)$ separating the dot from the ring and/or 
the potential well offset $V_{\rm QD}$-$V_{\rm QR}$, (see Fig. 1) one can considerably alter its optical, conducting and coherent properties.

In particular, we show that by such manipulations one can change:
 
a) the spin relaxation time $T_1$ of DRNs, used as spin qubits or spin memory devices, by orders of magnitude
 
b) the cross section for frequency selective optical absorption at infrared and microwave range from strong to negligible,

c) the conducting properties of an array of DRNs from highly conducting to insulating.

These features are mostly determined by the so called overlap factor (OF) (given by Eq. \ref{xi}) which reflects the shape and distribution of the wave functions and can be largely modified by the form of the confinement potential. Thus the microscopic properties of a DRN can be engineered on demand, depending on a particular application of the DRN.
 
The paper is organized as follows: In Sec. \ref{sec2} we present a general theoretical background that will be needed to study particular properties of DRNs. In the proceeding sections we demonstrate, by changing the parameters of the confinement potential, that we can control spin relaxation time (Sec. \ref{sec3}), optical absorption (Sec. \ref{sec4})
and transport properties (Sec. \ref{sec5}) of DRNs. The results are summarized in Sec. \ref{sec6}.

\section{Basic theoretical formulas}\label{sec2}

\begin{figure}[h]
\includegraphics[width=0.9\linewidth]{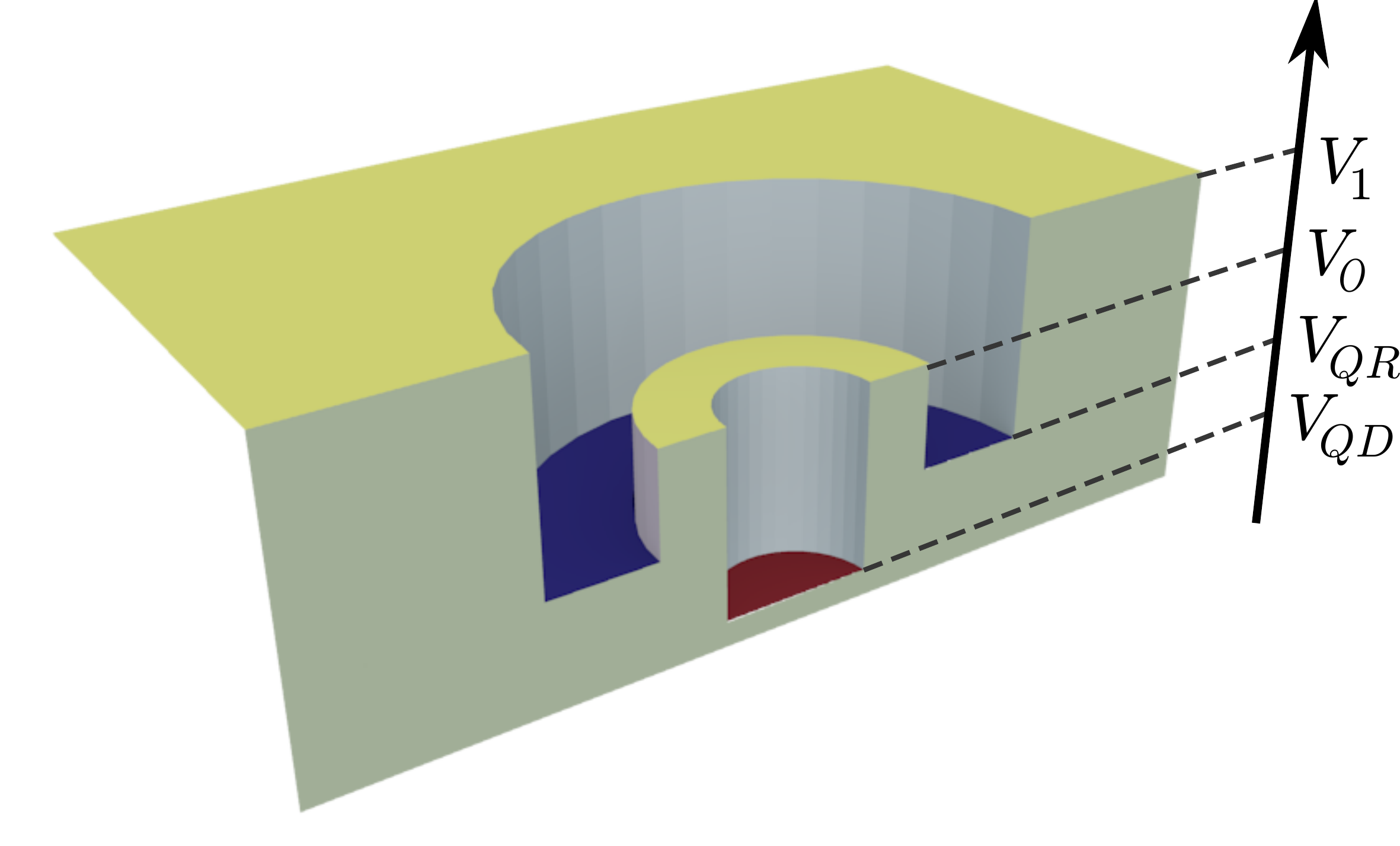}\vspace*{2mm}
\caption{(color online) Cross section of the confining potential of a DRN with marked bottom of the QD potential ($V_{\rm QD}$), bottom of the QR potential ($V_{\rm QR}$), top of the barrier potential ($V_0$) and the value of the potential outside the DRN ($V_1$).}
\label{fig1}
\end{figure}

We consider a 2D, circularly symmetric dot-ring nanostructure defined by a confinement potential $V(r)$. The DRN is composed of a QD surrounded by a QR and separated from the ring by a potential barrier $V_0$.
We assume that the barrier is sufficiently small to allow the electron tunnelings between the QD and QR. It assures that if we change the confining potential at low temperature the electron always occupies the ground state independently of the previous shape of the potential. A cross section of a DRN with explanations of symbols used throughout the text is presented in Fig.~\ref{fig1}. 

Such a confinement potential, which conserves the circular symmetry, can be obtained in many ways, e.g., using atomic force microscope to locally oxidize the surface of a sample,\cite{fuhrer} by self-assembly techniques [in particular by pulsed droplet epitaxy (PDE)],\cite{somaschini,mano} by the split gates,\cite{zhitenev} or by lithography.
In the calculations we assume cross sections of the potentials to have rectangular shapes. 
However, the results for more realistic potentials\cite{lis} differ only quantitatively, what will be presented elsewhere. The influence of magnetic field on such single and a few electron systems has been investigated in Ref. \onlinecite{peeters}.

\begin{figure}[h]
\includegraphics[width=0.46\linewidth]{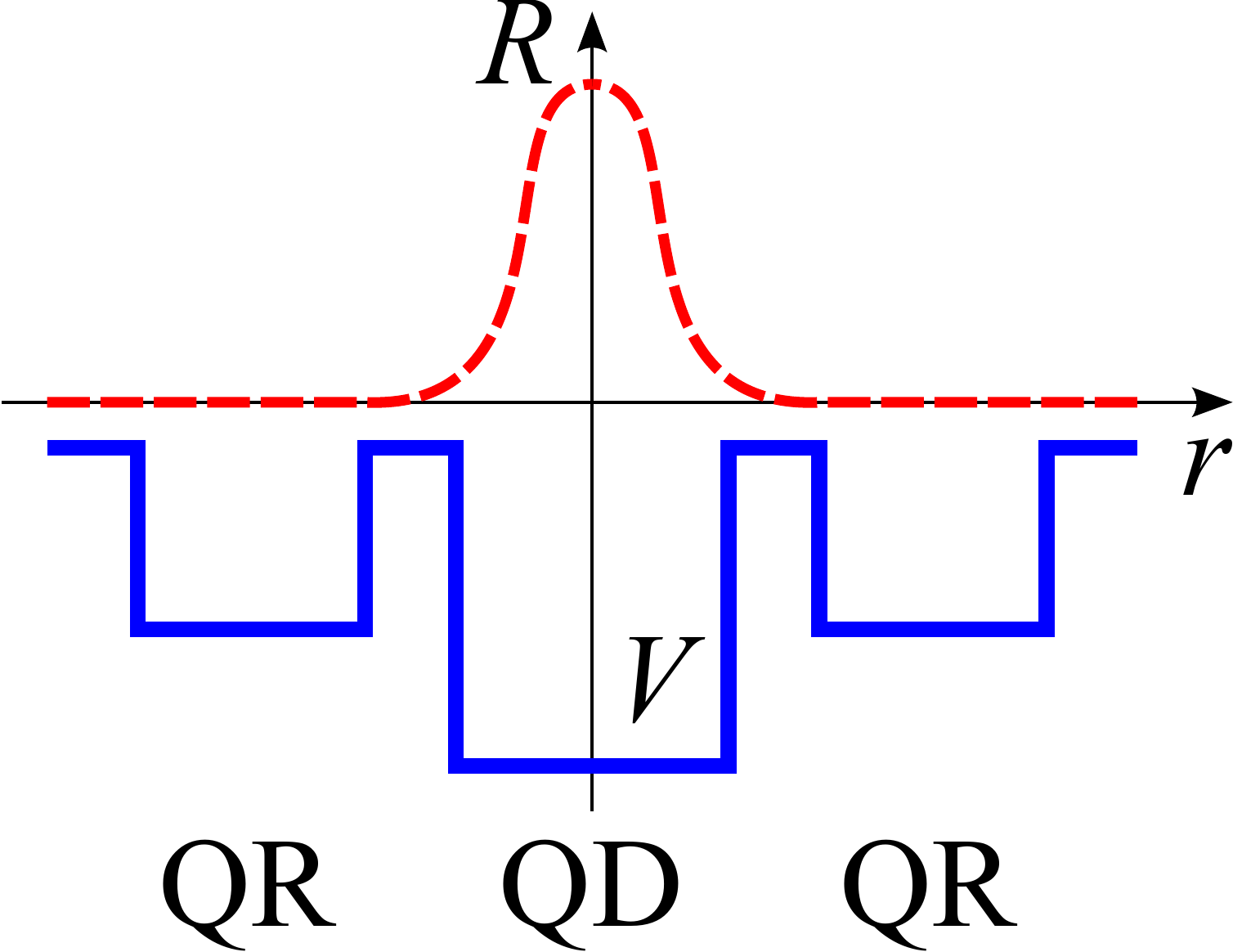}\vspace*{2mm}
\includegraphics[width=0.46\linewidth]{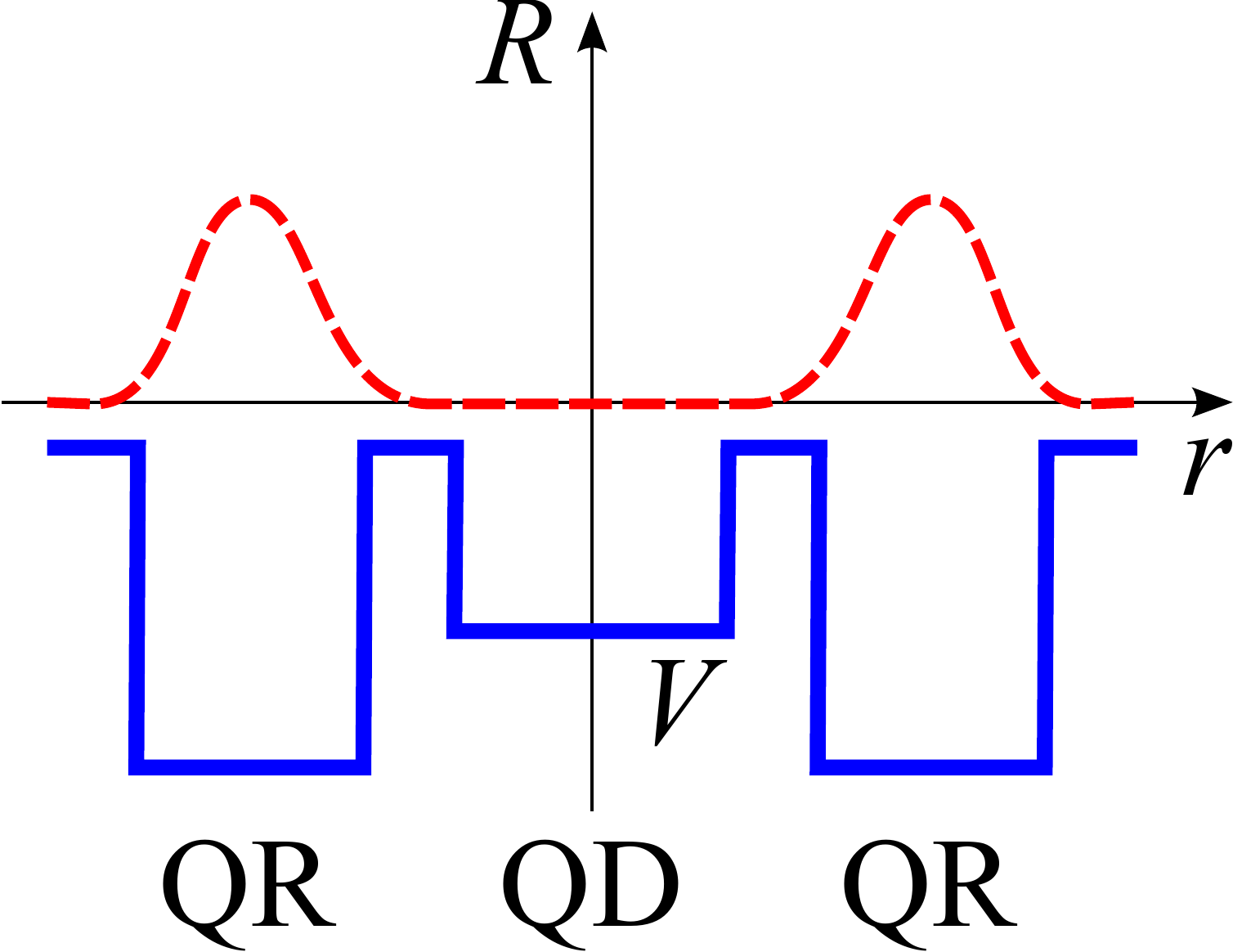}
\caption{(color online) Schematic illustration of the radial part of the ground state wave function (dashed red line) in the case when the QD potential is deeper than the QR potential (left panel) or {\em vice versa} (right panel). The cross section of the confining potential is represented by the solid blue line.}
\label{fig2}
\end{figure}

As it was already stated, the main advantage of DRN is the controllability of the shape of the electron wave functions. The main parameters that affect this shape are the relative positions of the bottoms of the QD and QR confining potentials and the size of the barrier between them. All these parameters can be tuned, e.g., by electrical gating. 
Roughly speaking, if the potential of the QD is much deeper than the potential of the QR, the electrons are located mainly in the QD and the effective size of the wave function is small. On the other hand, if the ring's potential is much deeper the electrons occupy only states in the QR and the wave function is much broader (Fig. \ref{fig2}). Moreover, by fine--tuning the confinement potential we are able to have, e.g., the ground state located in the QD, whereas the lowest excited state in the QR (or {\em vice versa}). This way we can easily control the OF and all the properties which depend on it. In Sections \ref{sec3} and \ref{sec4} we show how this feature can be exploited to control relaxation time and optical absorption.

For concreteness we assume that the radius of the DRN $r_0=40$ nm, $V_1=150$ meV and we set the zero potential energy at the level of $V_{\rm QR}$, i.e., the potential well offset is equal $V_{\rm QD}$. 
To be specific, our model calculations are performed for InGaAs systems (with $m^*= 0.067m_e$, $|g_s|=0.8$) for which many of the theoretical and experimental investigations have been done.

The single--electron Hamiltonian with the confinement potential $V(r)$ and in the presence of the in-plane magnetic field $B$ is written as
\begin{equation}
H = \frac{1}{2m^{*}} {\bf p}^2 + \frac{e\hbar}{2m^*} \mathbf{\hat{\sigma}} \cdot \mathbf{B}+V(r), 
\label{Hamiltonian_pr}
\end{equation} 
where $ m^{*}$ is the effective electron's mass.\\
The energy spectrum of $H$ consists of a set of discrete states $E_{nl}$ due to radial motion with radial quantum numbers $n=0,1,2,\ldots$, and rotational motion with angular momentum quantum numbers $l=0,\pm 1,\pm 2\ldots$.
The single particle wave function is of the form 
\begin{equation}
\Psi_{nl\sigma} = R_{nl}\left(r\right)\exp\left(i l \phi \right)\chi_{\sigma},
\label{eq_psi_nl}
\end{equation}
with the radial part $R_{nl}(r)$ and the spin part $\chi_{\sigma}$.
 
For many QDs the energy spectra and the wave functions can be calculated analytically,
for other structures have to be calculated numerically.
Having given the wave functions one can calculate the OF. It reflects the mutual distribution of radial parts of any two wave functions in the DRN, and is given by
\begin{equation}
\Xi_{n'l',nl} = \int_0^{\infty} R_{n'l'}^{*}R_{nl}r^2 dr,
\label{xi}
\end{equation}
where $n'l'$ and $nl$ are the quantum numbers of the two energy states involved in the process under investigation. In the following we consider DRN occupied by a {\it single} electron which couples to photonic (optical absorption) or phononic (spin relaxation) degrees of freedom. For such processes the selection rules allow the electron occupying the ground state to transit/scatter to the states with orbital numbers $l=\pm1$ only. Then, the relevant OF takes the form

\begin{equation}
\Xi_{00,nl}\equiv \Xi_{nl} = \int_0^{\infty} R_{00}^{*}R_{nl}r^2 dr,
\label{xi0}
\end{equation}
where $ R_{00}$ is the radial part of the ground state wave function (\ref{eq_psi_nl}) and $l=\pm 1$.
The second important quantity that depends on confinement and strongly affects absorption and relaxation is the energy gap between the orbital excited and the ground state 
\begin{equation}
\Delta_{nl}=E_{nl}-E_{00}. 
 \label{delta}
 \end{equation}
The calculated energy levels, modified by electrical gating, will be used in Section III to estimate the relaxation times for a set of DRNs.

\section{Spin relaxation times in dot--ring nanostructures}\label{sec3}

The spin of a {\it single} electron in circularly symmetric DRN placed in a static magnetic field $B$ with energy levels split by the Zeeman energy
 $\Delta _Z=g_s \mu_B B$ provides a natural system suitable as a memory device in spintronics and as a qubit in a quantum computer.\cite{Lossa} If $k_BT\ll\Delta _Z\ll\Delta _{01}$ then, the DRN can be well approximated as a two level system. The evident goal is to optimize material properties and nanostructure design to achieve long relaxation (and decoherence) times so that sufficient room is left for implementing protocols for spin manipulations and read out. 

In our model calculations we assume the in-plane magnetic field $B=1$ T (the in-plane orientation is favorable as it does not reduce the distance between the orbital states), thus the Zeeman splitting is equal to $\Delta_Z=0.046$ meV.

It was shown both theoretically \cite{Khaetskii,golo} and experimentally \cite{kroutvar, amasha, heiss} that the most important spin relaxation mechanism in magnetic fields of the order of a few Tesla is spin-orbit (SO) mediated spin-piezoelectric phonon interaction. The extensive discussion of the relaxation times in quantum dots and rings has been also given in Ref. \onlinecite{zkm}. In all these studies it was shown that relaxation times increase strongly with the decrease of the nanostructure radius, so we will not consider this aspect here. We rather focus on the change of $T_1$ governed by the change of the confinement potential that defines the investigated DRN structure.

The formula for the relaxation time $T_1$ governed by the Dresselhaus SO interaction is given by (for detailed derivation see Ref. \onlinecite{Khaetskii}):
\begin{equation}
\frac{1}{T_1}=\frac{\Delta_Z^5}{\eta}\left( \sum_{n,l} \frac{\Xi_{nl}^2}{\Delta_{nl}}\right)^2,
\label{t1_r}
\end{equation}
\begin{equation}
\eta=\frac{\hbar^5 }{\Lambda_p (2\pi)^4 (m^{*})^2 }.
\end{equation}
$\Lambda_p$ is the dimensionless constant depending on the strength of the effective spin-piezoelectric phonon coupling and the magnitude of SO interaction, $\Lambda_p=0.007$ for GaAs type systems.\cite{kroutvar,Khaetskii}
We have checked that for a single electron the relaxation time is determined by the SO coupling to (at most) two lowest excited orbital levels allowed by the selection rules, thus 
\begin{equation}
\frac{1}{T_1}=\frac{\Delta_z^5}{\eta}\left(\Gamma^{\rm 01} + \Gamma^{\rm 11}\right)^2,
\label{t1_dr}
\end{equation}
where
$$
\Gamma^{\rm 01}=\frac{\Xi^2_{01}}{\Delta_{01}},\ \ \Gamma^{\rm 11}=\frac{\Xi^2_{11}}{\Delta_{11}}.
$$
The quantities entering $T_1$ depend on the potential confining the electrons which determines the orbital energy spectrum, the shape of the orbital wave functions and therefore the OF.

\begin{figure}
\includegraphics[width=\linewidth]{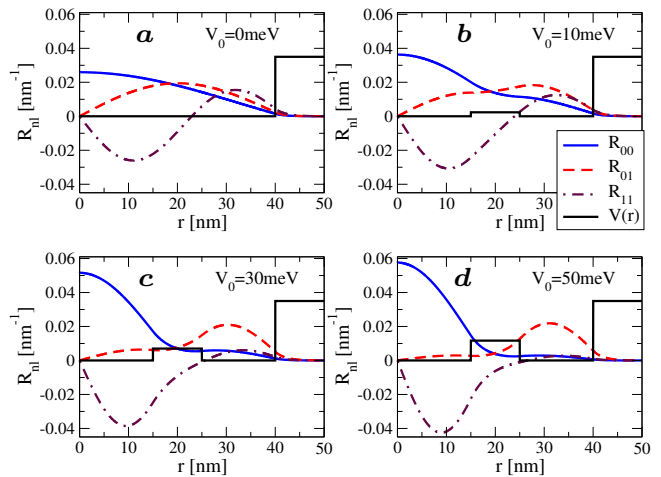}
\caption{(color online) The distribution of the wave functions of the three lowest energy states involved in spin relaxation for different values of $V_0$. $V_{\rm QD}=V_{\rm QR}=0$ meV and $V_1=150$ meV is assumed.}
\label{fig3}
\end{figure}

Let us first consider a limiting case when $V_{\rm QD}=V_0=0$. Then the DRN is a circular quantum dot with radius $r_0=40$ nm.
The calculated quantities are collected in Table \ref{table1} (first row) and the corresponding wave functions are presented in Fig. \ref{fig3}a. 

\begin{figure}
\includegraphics[width=\linewidth]{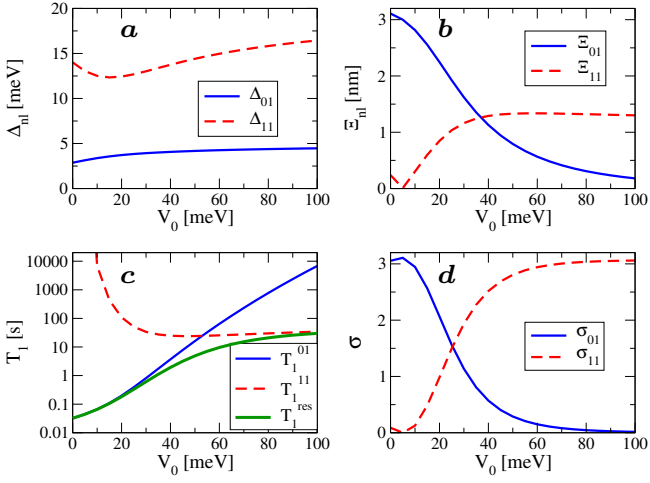}
\caption{(color online) The dependence of the orbital gaps $\Delta_{nl}$ (a), overlap factors $\Xi_{nl}$ (b), relaxation times (c) and photon absorption cross section (d) on the height of the potential $V_0$. Other parameters ($V_{\rm QD},\ V_{\rm QR}$ and $V_1$) are the same as in Fig. \ref{fig3}. The solid green line shows the overall relaxation time of the DRN, whereas blue and red lines represent individual relaxation times for phonon coupling to the $R_{01}$ and $R_{11}$ states, respectively. }
\label{fig4}
\end{figure}

\begin{table*}
\caption{\label{table1}The values of $\Delta_{nl}$, $\Xi_{nl}$, $T_1$ and $\sigma_{i,f}$ as a function of the height of the barrier $V_0$. The parameters are: $r_{\rm QD}$=15 nm, $r_{\rm barrier}$=10 nm, $r_{\rm QR}$=15 nm, $V_{\rm QD}=V_{\rm QR}$=0 meV, $V_1$=150 meV.}
\begin{ruledtabular}
\begin{tabular}{|c|c|c|c|c|c|c|c|}
\hline
 $V_0$& $\Delta_{01}$[meV] & $\Delta_{11}$[meV] & $\Xi_{01}$ [nm] &$ \Xi_{11}$ [nm]& $T_1$[s]& $\sigma_{01}$[nm$^2$] & $\sigma_{11}$[nm$^2$]\\ \hline
 0 & 3.14 & 13.0 & 3.11 &0.24&0.03 & 3.06 & 0.088\\ \hline
10 & 3.383 & 12.51 & 2.8&0.3 &0.066 & 2.94 & 0.13\\ \hline
30& 3.93 & 13 &1.62&1.16 & 0.61& 1.13 & 1.94 \\ \hline
50 & 4.18 & 14.4 &0.79 & 1.33& 4.89& 0.29 & 2.8 \\ \hline
75 & 4.35 & 15.6 & 0.35& 1.32&18.17 &0.06 &3 \\ \hline
\end{tabular}
\end{ruledtabular}
\end{table*}
The results show that in this case $T_1$ is entirely determined by the virtual excitation to the first excited orbital state and the reason of small $T_1$ is the relatively large $\Xi_{01}$. The second allowed excited state ($n=1$, $l=1$) lies too far in energy to let the electron scatter and the symmetry of the wave function (the dot-dashed line in Fig. \ref{fig3}a) causes small value of $\Xi_{11}$. Thus this state does not contribute. It is also seen in Fig. \ref{fig4}c where the relaxation times, calculated independently for transitions $00-01$ ($T_1^{01}$) and $00-11$ ($T_1^{11}$), are plotted. The resulting relaxation time of the nanostructure $T_1$ (solid green line in Fig. \ref{fig4}c) for $V_0=0$ overlaps with $T_1^{01}$.\\
When the barrier $V_0$ is present, the structure divides, forming a DRN and the wave function distribution changes drastically (Fig. \ref{fig3}b,c,d). In Figs. \ref{fig4}a and \ref{fig4}b we plotted $\Delta_{nl}$ and $\Xi_{nl}$ respectively as a function of $V_0$, the values are given in Table \ref{table1}.
Comparing these figures we see that, in contrast to $\Xi_{nl}$, $\Delta_{nl}$ is almost constant with $V_0$ and it is $\Xi_{nl}$ which determines $T_1$. Indeed, when $\Xi_{nl}$ increases then $T_1^{nl}$ decreases and {\it vice versa}.
For small $V_0$ the dominant contribution to relaxation is given by the state $E_{01}$ (Fig. \ref{fig4}c). With increasing $V_0$ the wave function of $E_{01}$ moves over to QR which results in a decrease of $\Xi_{01}$ (increase of $T_1^{01}$) with simultaneous increase of $\Xi_{11}$ (decrease of $T_1^{11}$).
For $V_0\approx 52$ meV the contributions to $T_1$ from $E_{01}$ and $E_{11}$ are equal and by further increasing the height of the barrier it becomes the {\it higher} excited state ($E_{11}$) that determines $T_1$ - a rather unusual situation. Similar phenomenon has been obtained by changing the shape of the QD by electrical gating.\cite{amasha}

Summarizing this part, we have shown that by changing the barrier height 
and shape one can change considerably the relaxation time of DRN by manipulating the orbital energy states and their wave functions.

\begin{figure}
\includegraphics[width=\linewidth]{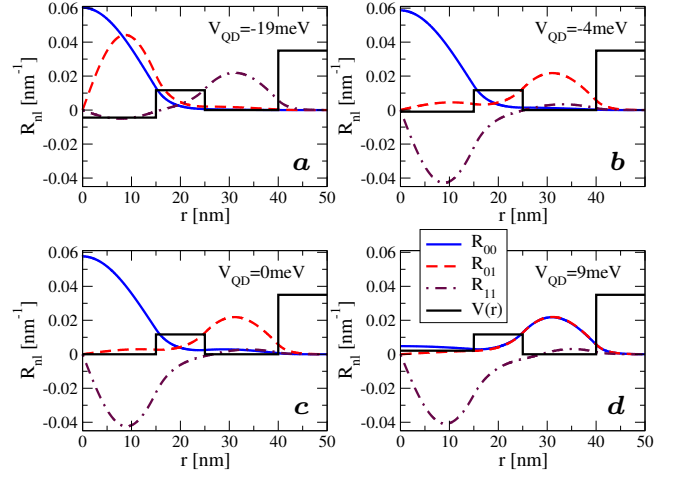}
\caption{(color online) The distribution of the wave functions of the first three energy states and for different values of $V_{\rm QD}$. $V_0=50$ meV and $V_1=150$ meV is assumed.}
\label{fig5}
\end{figure}

\begin{figure}
\includegraphics[width=\linewidth]{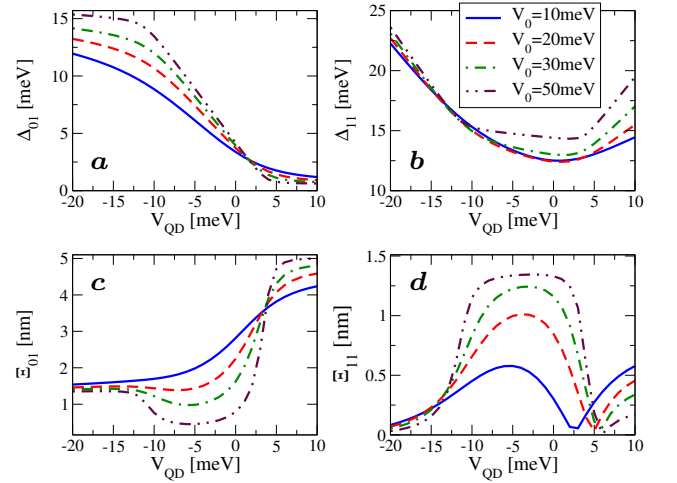}
\caption{(color online) Dependence of the orbital energy gap $\Delta_{nl}$ (panels a and b) and overlap factor $\Xi$ (panels c and d) to the first (panels a and c) and second (panels b and d) excited state as a function of $V_{\rm QD}$ for different values of $V_0$.}
\label{fig6}
\end{figure}

Similar considerations can be done for DRNs by changing, instead of the barrier height, the potential well offset $V_{\rm QD}$. Such manipulations can be easily done experimentally by the application of the gate potential below the QD. The wave functions of the three lowest states for different values of $V_{\rm QD}$ are shown in Fig. \ref{fig5}. One can see that changing the depth of the QD we can move individual wave functions between the QD and QR. By changing considerably the $V_{\rm QD}$ we can model the geometry of nanostructure from QD through DRN to QR. The possible applications of such features are given in Section \ref{sec5}.
The results of the calculations of relevant $\Delta$'s and $\Xi$'s are presented in Fig. \ref{fig6} and the corresponding relaxation times are given in Table \ref{table2}.

\begin{table*}
\caption{\label{table2}The values of $\Delta_{nl}$, $\Xi_{nl}$, $T_1$ and $\sigma_{i,f}$ as a function of the height of the barrier $V_0$. The parameters are : $r_{\rm QD}$=15 nm, $r_{\rm barrier}$=10 nm, $r_{\rm QR}$=15 nm, $V_{0}$=30 meV, $V_{\rm QR}$=0 meV, $V_1$=150 meV.}
\begin{ruledtabular}
\begin{tabular}{|c|c|c|c|c|c|c|c|}
\hline
 $V_{\rm QD}$& $\Delta_{01}$[meV] & $\Delta_{11}$[meV] & $\Xi_{01}$ [nm] &$ \Xi_{11}$ [nm]& $T_1$[s]& $\sigma_{01}$[nm$^2$] & $\sigma_{11}$[nm$^2$]\\ \hline
-19& 14.06 & 22.16 & 1.42 & 0.07 & 17.9 & 3.1 & 0.012\\ \hline
-10& 12  & 15  & 1.22 & 0.83 & 12.41 & 1.97 & 1.14\\ \hline
-4 & 7.35 & 13.5 & 1.02 & 1.24 & 5.56 & 0.84 & 2.28\\ \hline
0 & 3.93 & 13  & 1.62 & 1.16 & 0.61 & 1.14 & 1.94\\ \hline
4 & 1.38 & 13.45 & 3.88 & 0.37 & 0.03 & 2.30 & 0.20\\ \hline
9 & 0.78 & 16.38 & 4.8 & 0.30 & 0.004 & 1.97 & 0.16\\ \hline
\end{tabular}
\end{ruledtabular}
\end{table*}

One should stress that the important feature of such studies is not only the value of $T_1$ itself but the possibility to change it by external conditions which can be steered by electric fields. 
Besides these very long relaxation times have been obtained taking into account only SO mediated interaction with piezoelectric phonons.
 However other mechanisms of relaxation, (e.g., due to fluctuations of the electric and magnetic field, deformational phonons, multiphonon processes, and circuit noise) which we neglected in the above model calculations, can further limit the relaxation time.

In the next chapter we discuss another quantity which is determined by the DRN parameters, namely the intraband absorption of microwave and infrared radiation.

\section{Engineering optical absorption of DRNs}\label{sec4}

The cross section for photon absorption due to electron transition from the $i$-th bound state $E_i(n_i,l_i)$ to the $f$-th bound state $E_f(n_f,l_f)$ in the dipole approximation is given by the formula \cite{chakra,mila,bondar}
\begin{equation}
\sigma_{i,f} =\frac{16\pi^2\beta \hbar \omega \Xi_{i,f}^2 }{n_2} \delta(E_f-E_i -\hbar\omega)F_{\rm FD}(E_i,E_f)
\label{A}
\end{equation}
where $\beta=1/137$ is the fine structure constant, $l_f=l_i\pm1$, $n_2$ is the refractive index and
$F_{\rm FD}(E_i,E_f) \equiv f_{\rm FD}(E_i) - f_{\rm FD}(E_f)$, with $f_{\rm FD}(E)$ being the Fermi-Dirac distribution function. In these considerations we assume $B=0$ and $k_BT\ll\Delta _{01}$. Replacing the delta function by the Lorentzian function with half-width $\Gamma$ and neglecting the influence of temperature we obtain the maximum cross-section at the resonance frequency
\begin{equation}
\sigma_{i,f}^m =\frac{16\pi\beta \Xi_{i,f}^2 }{n_2 \Gamma} \Delta_{i,f},
\label{B}
\end{equation}

With the help of Eq. (\ref{B}) one can analyze the frequency selective absorption for a range of initial and final states. For concreteness we calculate the absorption coefficient from the orbital ground state $E_{00}$ to the first excited orbital state $E_{01}$ at frequency $\omega_{01}$ ($\hbar\omega_{01} \simeq \Delta_{01}$) and to the second excited state $E_{11}$ at frequency $\omega_{11}$ ($\hbar\omega_{11} \simeq \Delta_{11}$) and show how it can be modified for different DRNs.

\begin{figure}
\includegraphics[width=\linewidth]{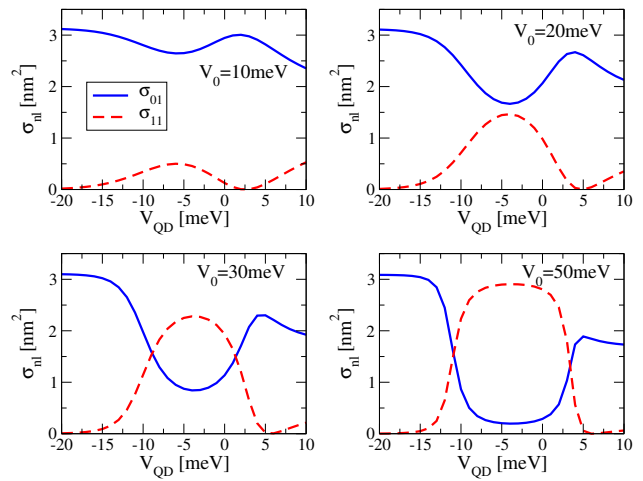}
\caption{Photon absorption cross section for transition from the ground state to the first (solid blue line) and the second (red dashed line) state as a function of $V_{\rm QD}$ for different values of $V_0$.}
\label{fig7}
\end{figure}

The absorption cross sections for the first two optical transitions as a function of the barrier height $V_0$ 

\begin{equation}
\sigma_{00,n1}^m = \sigma_{n1} = \frac{16\pi^2\beta\Delta_{n1}\Xi_{n1}^2}{n_2},\ (n=0,1)
\label{A1}
\end{equation}
(assuming $\Gamma = 1$ meV, $n_2=3.25$) are presented in Table I and in Fig. \ref{fig7}.

We see that for sufficiently large $V_0$ the absorption strongly depends on $V_{\rm QD}$. For $V_{\rm QD}$ located much below $V_{\rm QR}$ the wave functions of the ground and first excited state lie in the QD and that of the second excited state in the QR (see e.g., Fig. \ref{fig5}a). It results in high (small) absorption cross section $\sigma_{01}$ ( $\sigma_{11}$). With decreasing the potential well offset ($-5\:{\rm meV}<V_{\rm QD}<1\:{\rm meV}$) the distribution of the wave functions changes considerably resulting in $\sigma_{11}$ much bigger than $\sigma_{01}$. For further increasing $V_{\rm QD}$ the wave functions $R_{00}$, $R_{01}$ move over to QR and $R_{11}$ stays in QD (see e.g., Fig. \ref{fig5}d) which results in a reversed situation. For small $V_0$ these effects are much weaker.
 
These considerations allow us to engineer the DRNs according to their applications:

{\it i}) One can design DRNs to get most effective absorption required for efficient infrared and microwave photodetectors, or

{\it ii}) one can design DRNs with negligible absorption at $\hbar\omega_{01}$ or $\hbar\omega_{11}$, i.e., structures which will be transparent for the respective photon frequency. 

By changing $V_{\rm QD}$ one can smoothly move over from highly absorbing to almost transparent DRNs. 
The absorbed photon energy can be changed to large extent by changing the radius of DRN, the barrier height and the material (e.g., for the structure with $m^*= 0.04m_e$ it changes from microwaves to far infrared). 

\section{Conducting properties of arrays of DRNs}\label{sec5}

\begin{figure}[h]
\includegraphics[width=0.44\linewidth]{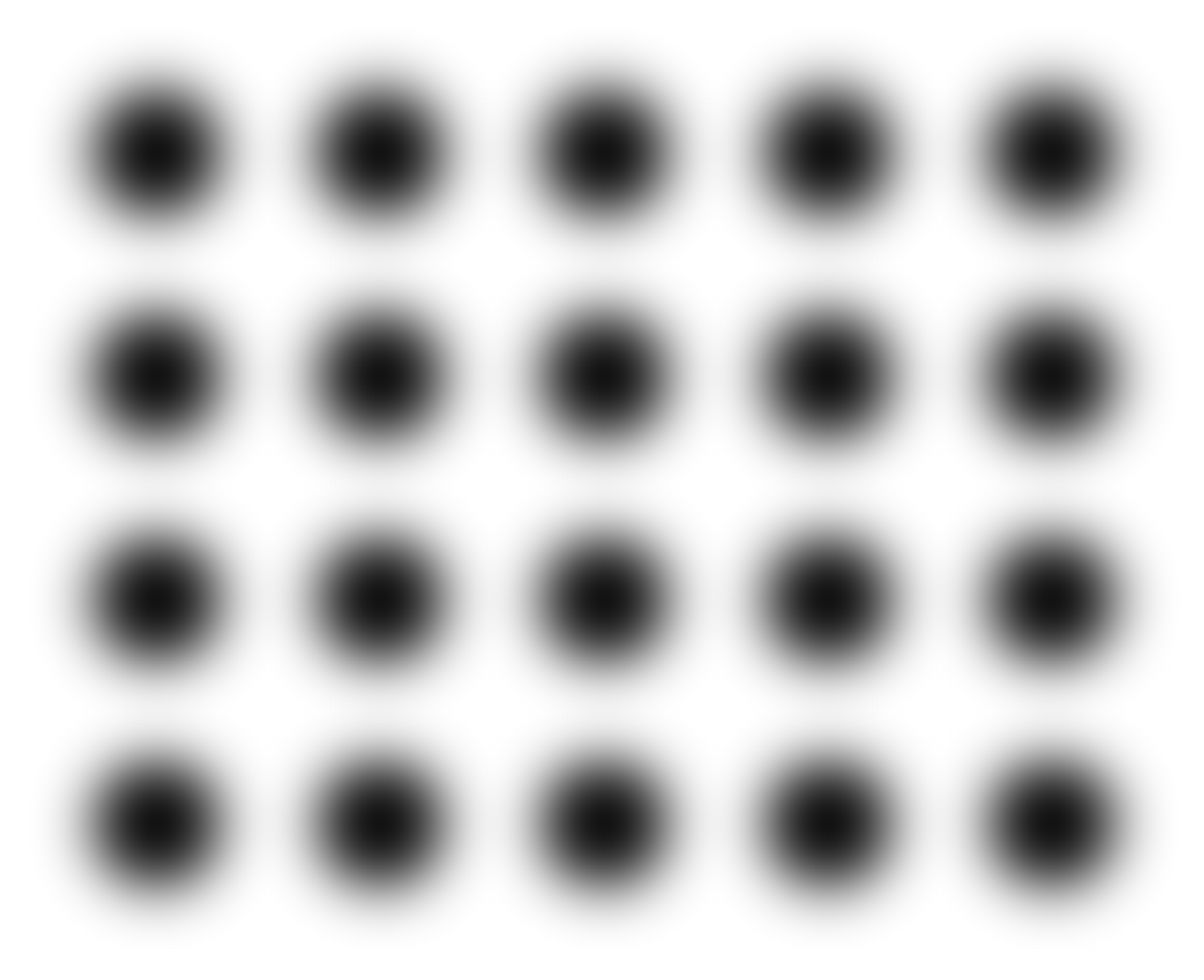}\vspace*{2mm}
\includegraphics[width=0.44\linewidth]{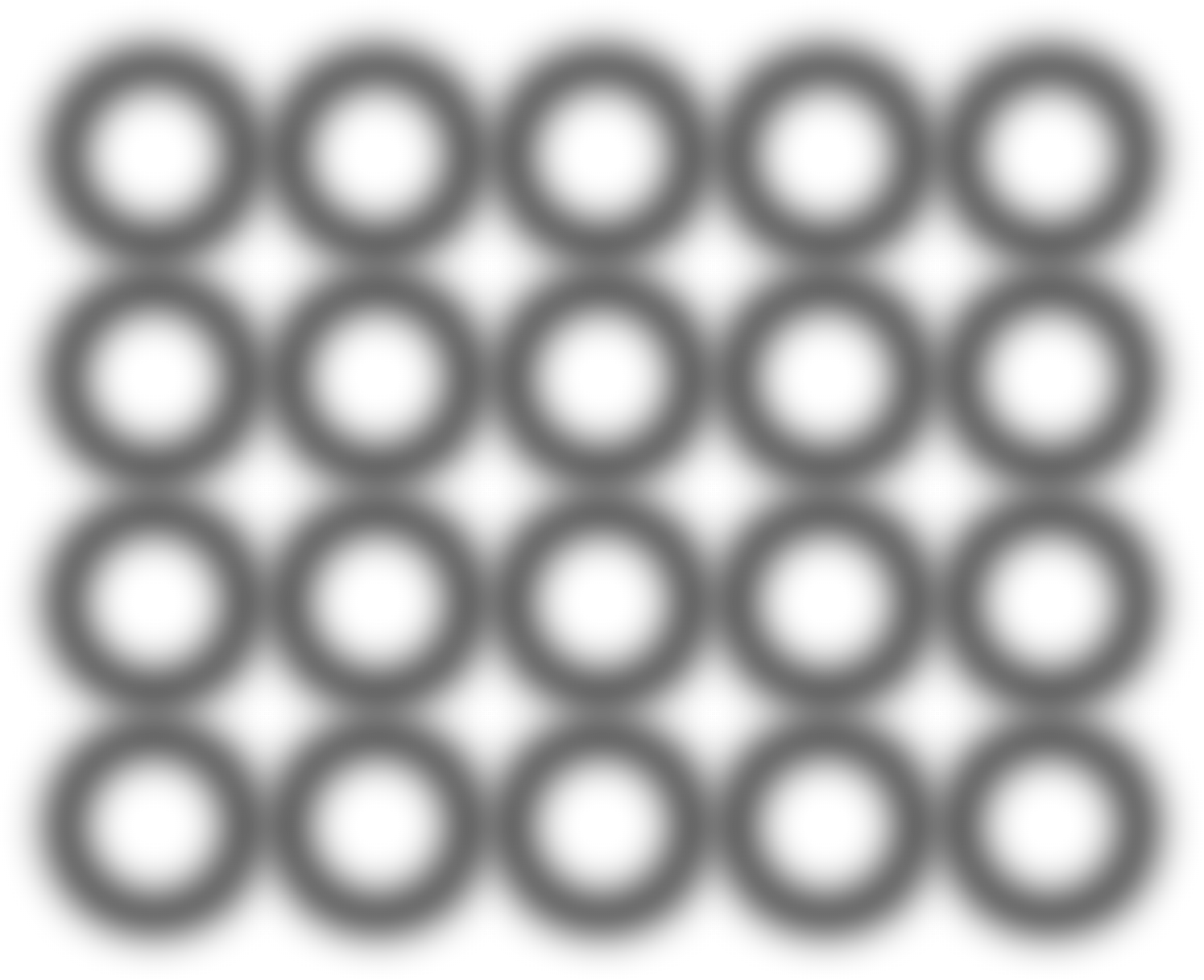}
\caption{Electron wave functions in an array of DRNs in the case when electron are located in the QDs (left panel) and in the QRs (right panel).}
\label{fig8}
\end{figure}

Apart from unique properties of a single DRN, interesting
behavior emerges when such structures are combined into a two-dimensional
array. If they are located sufficiently close to each other, electrons can 
tunnel from one DRN to another one, making a system that resembles 
a narrow band crystal. The tunneling rate depends on the overlap of the 
electron wave functions on adjoining structures. And since we are able to 
control the shape of the wave functions, we can control the overlap, and 
thereby manipulate the transport properties of the crystal-like structure. 
If the electron wave functions are located in the QDs, the overlap is effectively zero and
the system behaves like an insulator. On the other hand, when the wave 
functions are located in the QRs, the overlap is much larger which results in
metallic character. These two situations are illustrated in Fig. \ref{fig8}.

Since the electronic correlations are not taken into account in our approach, we do not expect a true metal--insulator transition. However, in real nanosystems the Coulomb blockade is present and when the number of electrons per DRN is exactly one (half-filled case) a Mott--type transition can occur.\cite{dassarma} Then, changing the gate voltage we would be able to induce the metal--insulator transition. 

If we assume that there is no tunneling between adjoining QDs, the system can be effectively described by a periodic--Anderson--like Hamiltonian\cite{PAM}
\begin{eqnarray}
{\cal H}&=&\sum_{\langle i,j\rangle\sigma}\left(t_{ij}-\epsilon_{\rm QR} \delta_{ij}\right)
c^{\dagger}_{i\sigma}c_{j\sigma}-\epsilon_{\rm QD}\sum_{i\sigma}f^{\dagger}_{i\sigma}f_{j\sigma} \\
&&+V\sum_{i\sigma}\left(f^{\dagger}_{i\sigma}c_{i\sigma}+{\rm H.c.}\right) +\frac{U_{\rm QD}}{2}\sum_in_{i}(n_{i}-1), \nonumber
\end{eqnarray}
where $c^{\dagger}_{i\sigma}$ ($f^{\dagger}_{i\sigma}$) creates an electron in $i$-th QR (QD), $t_{ij}$ is the transfer energy between adjoining QRs, $V$ describes hybridization between states in QD and QR at the same site. In accordance with the Coulomb blockade picture the interaction between electrons in a given QD can be parametrized by a capacitive charging energy $U_{\rm QD}=e^2/C$. Coulomb interaction in QRs is neglected. The difference of the atomic levels $\epsilon_{\rm QD}-\epsilon_{\rm QR}$ can be controlled by the voltage applied to the QD's gates.

There is also another possibility to induce the metal--insulator transition. When a random gate voltage is applied to DRNs an on--diagonal disorder is introduced into the array and the Anderson localization is expected.

\begin{figure}[h]
\includegraphics[width=0.45\textwidth]{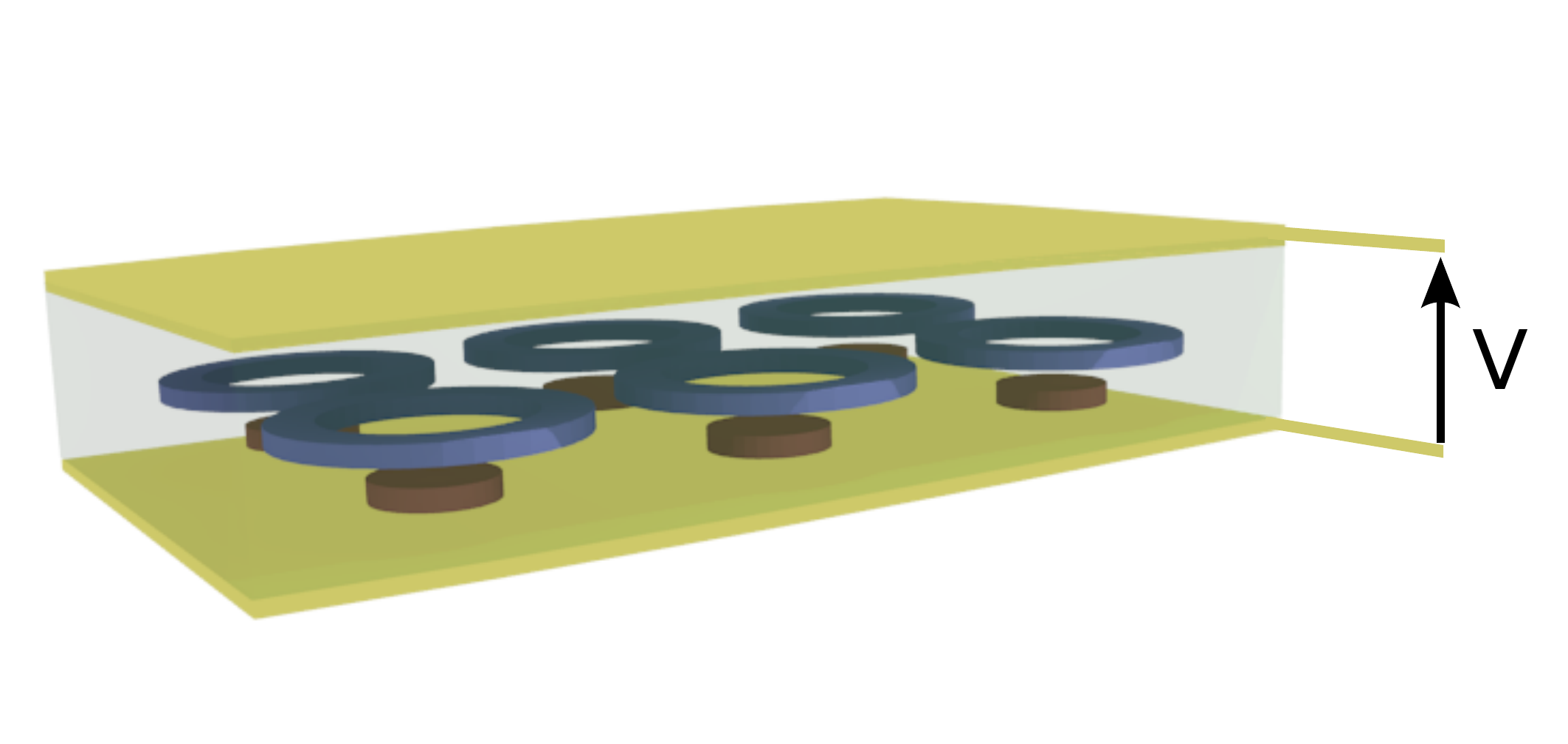}
\caption{(color online) Example of a spatial configuration of QR and QD that allows for controlling $V_{\rm QR}$ and $V_{\rm QD}$ by an external electrostatic field.}
\label{fig9}
\end{figure}

Controlling individual DRNs in an array may be involved since it requires to supply voltage to every gate. However, if we are interested only in global properties of such array, there is no need to control individual DRN's. In order to be able to control all the DRNs in the same way, i.e, to force electrons to occupy QDs or QRs in all nanostructures, we can place the QDs in one layer and the QRs in another one, located above (or below) the first layer. Then, the difference $\epsilon_{\rm QD}-\epsilon_{\rm QR}$ is proportional to the strength of electric field applied to the whole array perpendicularly to the layers. Such configuration is shown in Fig. \ref{fig9}. Of course, since in this case QDs and QRs are vertically separated, one can use sufficiently large QDs instead of QRs.
 
A similar approach has been proposed by Ugajin,\cite{Ugajin} where an array of coupled QDs was subjected to an external electric field. The QDs were of a convex shape and the applied field moved electrons into the convex portion what led to larger barrier width between adjacent dots and smaller transfer energy. 

\begin{figure}[h]
\includegraphics[width=0.6\linewidth]{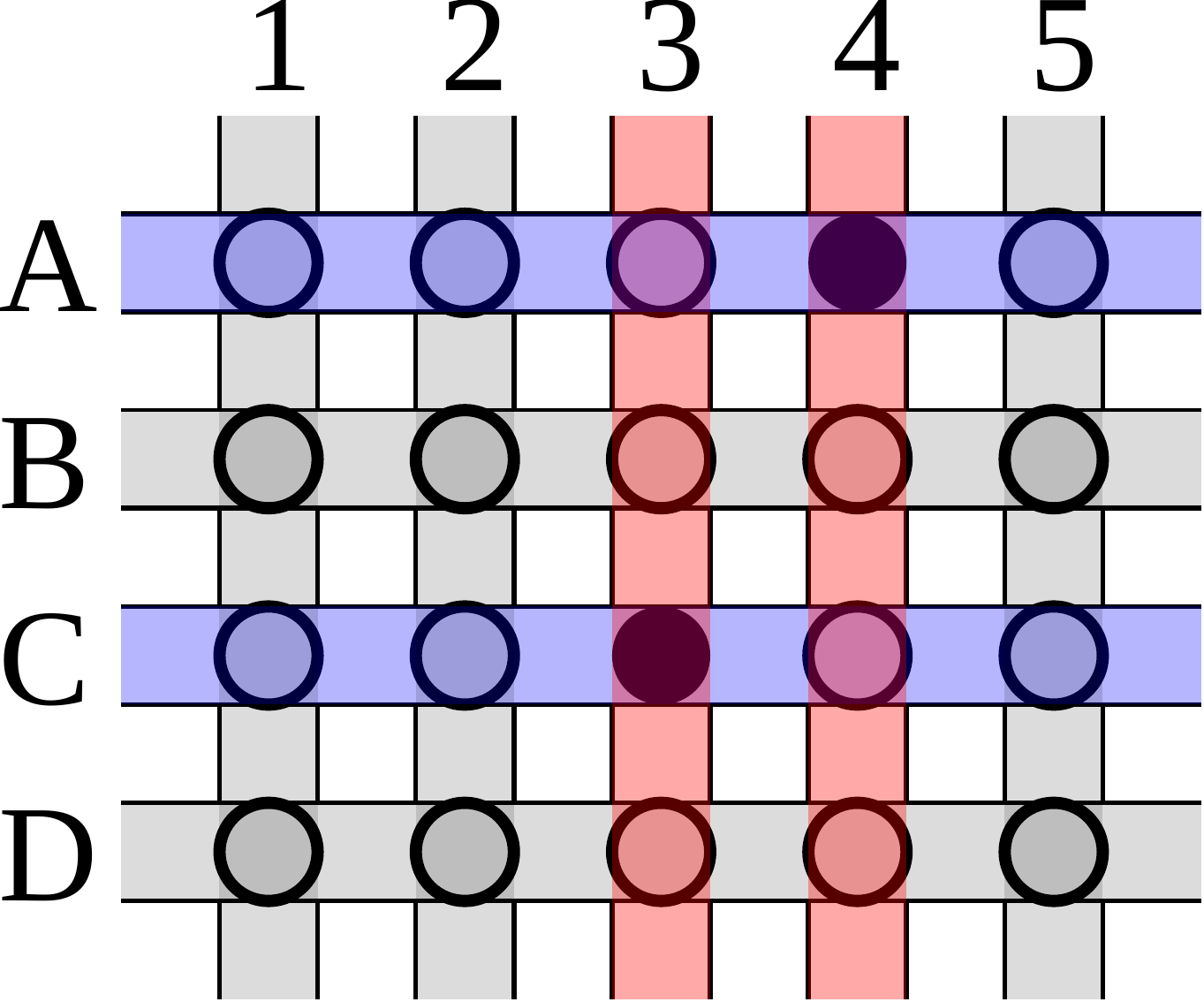}
\caption{(color online) Schematic illustration of the setup allowing for controlling individual DRNs. Stripes along the $x$ axis located above the DRNs are labeled with letters, whereas the stripes along the $y$ axis located below the DRNs are labeled with numbers. The full black circles indicate DRNs where the difference $\epsilon_{\rm QD}-\epsilon_{\rm QR}$ is largest when voltage is applied to pairs of stripes A4 and C3.}
\label{fig10}
\end{figure}

In arrays of DRNs with QRs and QDs located at different positions in the direction perpendicular to the array we can also control individual DRNs. It can be done in a way similar to addressing computer memory. Let us assume the array is in the $xy$ plane with the rows and columns of DRNs parallel to the axis. Then, we attach metallic stripes of width comparable to the diameter of the QDs above and below each row and column in such a way that the stripes above the array are parallel to, let say, the $x$ axis and the stripes below are parallel to the $y$ axes. Applying voltage to a given pair of stripes (one above the array and one below) produces the strongest electric field at the crossing and the nanostructure located there would be mostly affected. The situation is presented in Fig. \ref{fig10}. 

\section{Summary}\label{sec6}

The ability to control the quantum state of a single electron is at the heart of many developments, e.g., in spintronics and quantum computing. In contrast to real atoms, quantum nanostructures allow flexible control over the confinement potential which gives rise to wave function engineering.

We have shown that by manipulating the confinement parameters we can alter the overlap of the electron wave functions so that the transition probability will be enhanced or suppressed on demand. We performed systematic studies of the influence of such manipulations on relaxation times, optical absorption and conducting properties of dot-ring nanostructures. Thus the basic issues of quantum mechanics can be explored to design new semiconductor devices in which specific properties can be optimized. 

The wavelength range of the absorption spectra may be largely expanded from microwaves to infra--red by utilizing DRNs of different size and different material. 
The macroscopic variables such as the relaxation time, optical absorption or conductivity can be modified by changing on demand the microscopic features of the nanosystem such as the shape and distribution 
of the wave functions. 
Combined quantum structures are highly relevant to new technologies in which the control and manipulations of electron spin and wave functions play an important role. To name a few possible applications, one can imagine tunable or switchable microwave waveguides build with the help of arrays of DRNs with variable optical properties or a single electron transistor based on a single DRN coupled to source and drain leads.

The results indicate a novel opportunity to tune the performance of nanostructures and to optimize their specific properties by means of sophisticated structural design.

\acknowledgments
M.M.M. and M.K. acknowledge support from the Foundation for Polish Science under the ``TEAM'' program for the years 2011-2014. E.Z. acknowledges support from the Ministry of Science and Higher Education (Poland) under Grant No. N N202\:052940. The authors thank Jerzy Wr\'obel and Bart{\l}omiej Szafran for valuable discussions.

\end{document}